# Mono/dual-polarization refractive-index biosensors with enhanced sensitivity based on annular photonic crystals

**Liyong Jiang**[1,2,*], **Hong Wu**[1,3], **We Zhang**[1], and **Xiangyin Li**[1]

[1] Nanophotonic Laboratory, Department of Physics, Nanjing University of Science and Technology, Nanjing 210094, China

[2] Centre for Disruptive Photonic Technologies, School of Physical and Mathematical Sciences, Nanyang Technological University, 1 Nanyang Walk, Blk 5 Level 3, Singapore 637371

[3] School of Electrical and Computer Engineering, Georgia Institute of Technology, Atlanta, Georgia 30332-0250, USA

*corresponding author: jly@mail.njust.edu.cn

**Abstract:** To promote the development of two-dimensional (2D) photonic crystals (PCs) based refractive-index (RI) biosensors, there is an urgent requirement of an effective approach to improve the RI sensitivity of 2D PCs (usually less than 500 nm/RIU). In this work, the photonic band gap (PBG) feature and the corresponding RI sensitivity of the air-ring type 2D annular PCs (APCs) have been studied in detail. Such type of 2D PCs can easily and apparently improve the RI sensitivity in comparison with conventional air-hole type 2D PCs that have been widely studied in previous works. This is because the APCs can naturally exhibit suppressed up edge of PBG that can strongly affect the final RI sensitivity. In general, an enhanced sensing performance of as high as up to 2~3 times RI sensitivity can be obtained from pure 2D APCs. Such high RI sensitivity is also available in three typical waveguides developed from pure 2D APCs. Furthermore, a new conception of dual-polarization RI biosensors has been proposed by defining the RI sensitivity as the difference of crossing relative sensitivity of up-edge PBG between different polarizations. Based on pure 2D APCs, a dual-polarization RI biosensor with high RI sensitivity of ~1190 nm/RIU in infrared range was also presented as an example.

## 1. Introduction

Refractive-index (RI) biosensors are a representative type of label-free optical biosensors which have attractive applications in biomedical research, healthcare, pharmaceuticals, environmental monitoring and so on [1]. It is well-known that the most important two parameters to evaluate the performance of biosensors are sensitivity and detection limit (DL). In most cases, the sensitivity of an optical RI-based biosensor can be roughly defined as the shift of optical signal (such as wavelength, intensity or phase) in response to the change of RI in analyte. For instance, if the wavelength shift is considered the units of sensitivity can be usually written as nm/RIU, where RIU is the refractive index units. Accordingly, the DL is defined as the minimum resolvable RI change (in units of RIU) that can be detected by the sensor and it can be written as $DL = \delta/S$, where $S$ is the sensitivity and $\delta$ is the signal resolution of the optical spectrum analyzer corresponding to wavelength, intensity or phase. DL in units of RIU enables a rough comparison of the sensing capability among different optical technologies and structures. Sometimes, in many biomedical experiments, it is difficult or even impossible to determine the RI change in analyte. In these cases, the DL is usually defined as the surface total mass (in units of pg), the surface mass density (in units of $pg/mm^2$), or the sample concentration (in units of ng/mL or molarity).

In the past decade, RI-based biosensor development has been a fascinating and fast-paced area. The typical RI-based biosensor includes: (1) surface plasmon resonance (SPR) biosensors [2, 3]; (2) interferometer-based biosensors [4]; (3) optical waveguide based biosensors [5]; (4) optical ring resonator based biosensors [6]; (5) optical fiber based biosensors [7]; and (6) photonic crystals (PCs) based biosensors [8-20]. Note that it is common that an optical label-free sensor involves two or more optical structures mentioned above to enhance its sensing performance. Among these RI-based biosensors, the SPR biosensors were first studied and have been commercially applied (such as Biacore 2000). The SPR biosensors usually can show very high RI sensitivity (about several thousand nm/RIU) and much low DL ($10^{-5}$-$10^{-8}$ RIU) [1]. Recently, as a representative one of the new types of biosensors, PCs have gathered more and more interests from the researchers since they own some attractive features. One crucial feature is the strong field confinement in the photonic band gap (PBG) that can support strong light-matter interactions with a substance of interest. This feature is quite useful when sensing purpose of small amount or low density of analytes is proposed. Another crucial feature of PCs for bio-sensing application is their compact structure and natural micro channels that make them to be an ideal platform for microfluidic or optofluidic lab-on-chip technologies [18, 21]. The PCs-based biosensors can be further divided into four sub types, i.e., one-dimensional (1D) PCs biosensors [9, 10], two-dimensional (2D) PCs biosensors [11-18], three-dimensional (3D) PCs biosensors [19], and photonic crystal fiber biosensors [20]. However, as compared to SPR biosensors, the RI sensitivity and DL of PCs biosensors are still not satisfactory. In particular, 2D PCs biosensors have been paid much attention due to their relatively convenient in fabrication and compact structure for integration purpose, but their theoretical RI sensitivity is usually less than 500 nm/RIU with DL ($10^{-2}$-$10^{-5}$ RIU) [11-18] at near-infrared area.

The purpose of this paper is to report varies of new 2D PCs biosensors with obviously enhanced RI sensitivity based on 2D annular PCs (APCs). APCs were firstly reported by Kurt *et al.* [22] in 2005 and they are usually constructed by air rings embedded in a dielectric background or dielectric rings embedded in an air background. Compared with conventional air-hole or dielectric-rod type PCs, the most important feature of APCs is their natural capacity of well supporting both transverse electric (TE) and transverse magnetic (TM) polarization modes. This feature brings us more freedom to design kinds of polarization-insensitive devices [23-25]. Alternatively, APCs have also been used in RI sensing application. For instance, Säynätjoki *et al.* [26] have first mentioned the possibility of enhanced RI sensitivity by using 2D APCs instead of conventional 2D PCs. However, they did not give systematical investigations about it. Quite recently, Kurt *et al.* [27] reported a novel structure of RI biosensors by replacing one line of dielectric rods at the output surface of 2D PCs by one line of dielectric rings, which means a 1D dielectric-ring type APC. Their study shows that 1D dielectric-ring type APC can well support surface modes for bio-sensing application. Different from previous work, the present work will systematically investigate kinds of 2D air-ring type APCs based biosensors and show their enhanced RI sensitivity compared with present air-hole type 2D PCs biosensors. Moreover, we will present a new conception of dual-polarization 2D PCs biosensors based on pure APCs.

## 2. Mono–polarization 2D APCs biosensors

### 2.1 Enhanced RI Sensitivity in pure 2D APCs

In this section, we will discuss the enhanced RI sensitivity in pure 2D APCs. Before such discussion, it is better to firstly study the sensing properties of pure PCs for further comparison. It is well-known that the most important feature of PCs is the existence of PBG, during which the transmission of electromagnetic waves is forbidden. In general, the fundamental physical mechanism of pure PCs-based RI biosensors is the shift of PBG edge when RI of surrounding media is changed. Figure 1(a) shows the PBG response of TE-polarization mode to surrounding-RI change for conventional triangular-lattice hole-type Silicon ($n_{Si}$ = 12) 2D PCs with different air-hole radius. It can be found that for a specific air-hole radius, both up and bottom frequency edges will gradually decrease as surrounding RI increases, while the up frequency edge of PBG becomes more sensitive to the change of RI. As air-hole radius increases, which means the filling factor of air holes (defined as $2\pi R^2/\sqrt{3}a^2$) increases, the up frequency edge of PBG will show a blue-shift response. These behaviors can be simply explained by the band-gap theory. For air-hole PCs, the up edge and bottom edge of a PBG usually correspond to the edge of air (with lower RI) band and dielectric (with higher RI) band respectively. Since the air bands are located at higher frequency range than dielectric bands, the PCs with larger filling factor of air holes will correspond to higher up frequency edge of PBG while the PCs with air holes filled by larger surrounding-RI media will draw the up frequency edge of PBG to lower frequency ranges. At the same time, if the RI of surrounding media reaches the maximum limitation of the formation of PBG (which is around $n_{Si}/1.5$), the PBG will disappear and the sensing application will be unavailable. As a result, the up frequency edge shift can be used to define the RI sensitivity of pure PCs while the detection range of RI change is dominated by the existence of PBG. Figure 1(b) shows the corresponding response of up-edge sensitivity of PBG to surrounding-RI change.

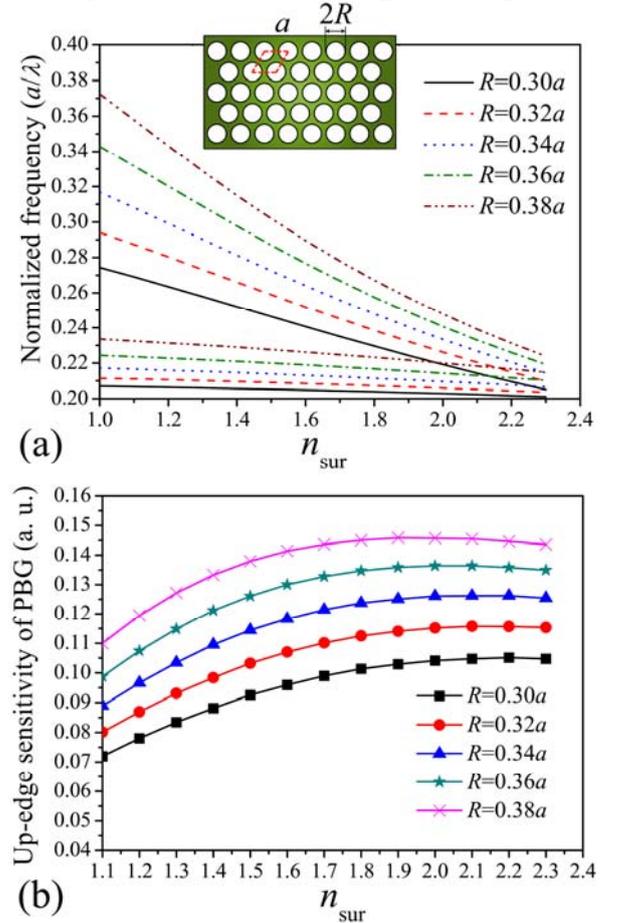

Fig. 1. (a) The PBG response to surrounding-RI change for conventional hole-type Silicon 2D PCs (see inset) with different air-hole radius $R$ ($a$ is the lattice constant). For each radius, the PBG range is represented by the distance between up and bottom frequency edges. (b) The corresponding response of up-edge sensitivity of PBG to surrounding-RI change.

In particular, we define the relative RI sensitivity as:

$$S_{(i+1)} = 1/\omega_{\text{up-edge}}^{n_{\text{sur}(i+1)}} - 1/\omega_{\text{up-edge}}^{n_{\text{sur}(i)}} \quad (1)$$

Here, $\omega_{\text{up-edge}}^{n_{\text{sur}(i)}}$ represents the up-edge frequency of PBG when the $i$-th surrounding RI $n_{\text{sur}(i)}$ is considered. It can be observed from Fig. 1(b) that the relative RI sensitivity will gradually increase (due to lower up-edge frequency of PBG) and finally reach at a steady level or only slightly drop down (due to the close of PBG) as the surrounding RI is linearly increased. Meanwhile, with the same surrounding RI, larger radius of air holes will show higher relative RI sensitivity. It is easy to make a conclusion that for conventional triangular-lattice hole-type PCs, a relatively large filling factor of air holes is recommended to get relatively high RI sensitivity.

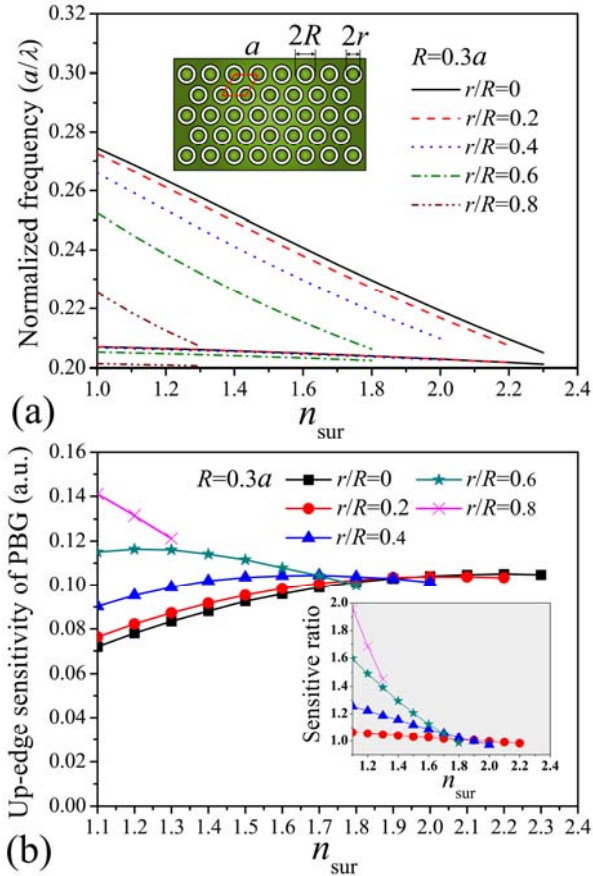

Fig. 2. (a) The PBG response to surrounding-RI change for pure 2D APCs (see inset) with fixed outer radius $R=0.3a$ and different inner/outer radius ratio $r/R$ of air rings. (b) The corresponding response of up-edge sensitivity of PBG to surrounding-RI change. The inset shows the sensitive ratio between different 2D APCs ($R=0.3$ while $r/R$ varies) and conventional hole-type 2D PCs when $R=0.3a$.

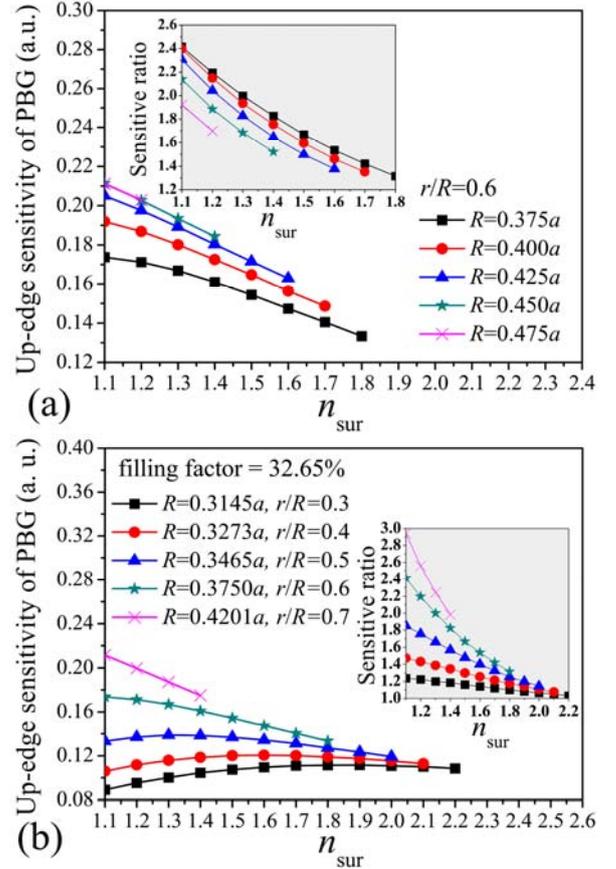

Fig. 3. (a) For pure 2D APCs with fixed inner/outer radius ratio $r/R=0.6$ and different outer radius $R$ of air rings, the corresponding response of up-edge sensitivity of PBG to surrounding-RI change. The inset shows the sensitive ratio between different 2D APCs ($r/R=0.6$ while $R$ varies) and conventional hole-type 2D PCs when the same $R$ is considered. (b) For pure 2D APCs (see inset) with fixed filling factor (32.65%) and different outer and inner radius of air rings, the corresponding response of up-edge sensitivity of PBG to surrounding-RI change. The inset shows the sensitive ratio between different 2D APCs (filling factor is 32.65% while both $R$ and $r/R$ vary) and conventional hole-type 2D PCs when the same filling factor 32.65% (i.e., $R=0.3a$) is considered.

Now let's turn our attentions to pure 2D APCs. We consider an air-ring type of 2D APCs as shown in the inset of Fig. 2(a). The outer and inner radius of air rings is represented by $R$ and $r$ respectively. As shown in Fig. 2(a), when $R$ is fixed to be $0.3a$ the APC with lager inner/outer radius ratio $r/R$ will show lower up-edge frequency of PBG. Such behavior is obviously due to the decrease of filling factor of air rings (defined as $2\pi R^2[1-(r/R)^2]/\sqrt{3}a^2$) according to above-mentioned PBG theory. The corresponding response of up-edge sensitivity of PBG to surrounding-RI change is plotted in Fig. 2(b). In comparison with conventional PCs ($r/R=0$), the detection range of RI change for APCs will gradually shrink as $r/R$ increases. This is due to the earlier close of PBG when $r/R$ increases as shown in Fig. 2(a). However, the APCs can show higher relative RI sensitivity during full detection range of RI change or in most detection range before a crossing with the curve of conventional PCs. Such crossing is also attributed to the unavoidable close of PBG. To see more detail about the enhanced RI sensitivity, the sensitivity ratio between APCs and conventional PCs ($r/R=0$) was calculated and shown in the inset of Fig. 2(b). Compared with conventional PCs, an enhanced RI sensitivity as high as around two times can be obtained when $r/R=0.8$. It is believed that such enhanced RI sensitivity is attributed to the suppressed up frequency edge of PBG [seen in Fig. 2(b)] that is in fact due to the phase contribution of additional dielectric rods in air holes.

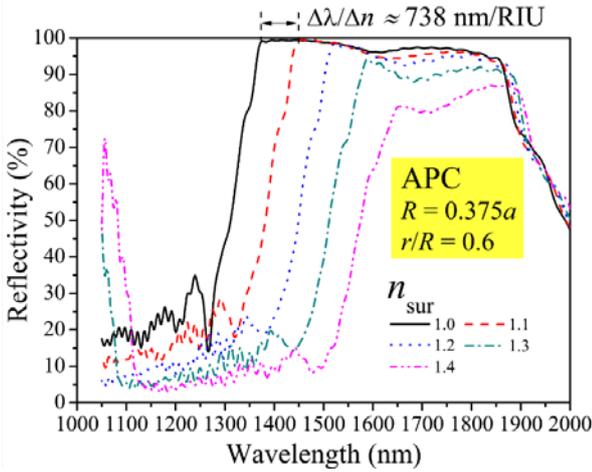

Fig. 4. Typical response of reflection spectra to surrounding-RI change for the pure APC when $R=0.375a$ and $r/R=0.6$.

Similar studies were also systematically executed when a fixed inner/outer radius ratio $r/R$ and a fixed filling factor of air rings were considered. As shown in Fig. 3(a), when the inner/outer radius ratio $r/R$ is fixed to be 0.6, APCs with different outer radius show very similar response of up-edge sensitivity of PBG to surrounding-RI change. The APCs with larger outer radius will show higher RI sensitivity but smaller detection range of RI change. Moreover, compared with conventional PCs with the same outer radius, APCs with larger outer radius will show lower sensitivity ratio. As shown in Fig. 3(b), when the filling factor of air rings is fixed to be 32.65% (in case of $R=0.375a$ and $r/R=0.6$), the same rule can be observed for both up-edge sensitivity and sensitivity ratio (see inset), that is APCs with larger outer radius will show higher RI sensitivity and sensitivity ratio. Compared with conventional PCs with the same filling factor (in case of $R=0.3a$ and $r/R=0$), an enhanced RI sensitivity as high as around three times can be obtained when APCs with $R=0.4201a$ and $r/R=0.7$ are considered. Such enhanced RI sensitivity can apparently improve the sensing capacity of PCs and make such type of RI-based biosensor more useful in practical application. Considering the requirements of both high sensitivity ratio and relatively large detection range of RI change, in section 2 we only consider the APCs structure with $R=0.375a$ and $r/R=0.6$ as an example to show kinds of APCs-based biosensor designs working only for TE polarization.

For instance, if we just use pure 2D APCs with $R=0.375a$ and $r/R=0.6$ as biosensor, we can make a transmission simulation with the help of finite-different time-domain (FDTD) method. Figure 4 gives a typical reflection spectrum of such APC when the lattice constant $a$ is assumed to be 420 nm. Such APC can work in near-infrared area with wavelength sensitivity of ~738 nm/RIU.

## 2.2 Enhanced RI Sensitivity in 2D APCs-based waveguides

Besides pure PCs, another widely used PCs-based biosensor structure is photonic crystal waveguides (PCWs). The most characteristic spectral feature created by the PCWs is a transmission cut-off near the PBG. In this section, we will further investigate the RI Sensitivity in kinds of 2D APCs-based waveguides. In particular, we will present three typical structures of PCWs as examples.

The first typical structure of PCWs is PCW-W1, which is constructed by removing one line of air holes in pure PCs along the propagating direction of electromagnetic waves. Based on such conception, a typical annular PCW-W1 (APCW-W1) structure and its project band diagram are plotted in Fig. 5(a). In similar with conventional PCW-W1, the APCW-W1 can show two fundamental TE-polarization modes, that is even mode and odd mode respectively. Both modes can be well confined in the line defect due to the PBG effect. When the RI of surrounding media varies, we only need to focus our attentions to the shift of up frequency edge of odd mode to investigate its RI sensing response. Figure 5(b) presents the up-edge sensitivity of odd mode for both APCW-W1 and conventional PCW-W1 with the same filling factor 32.65%. In good agreement with our prediction, APCW-W1 can support about two times higher RI sensitivity than conventional PCW-W1. The typical transmission spectra of such APCW-W1 were also calculated by FDTD method by assuming the lattice constant $a$ to be 420 nm. A wavelength sensitivity of ~511 nm/RIU can be obtained when RI changes.

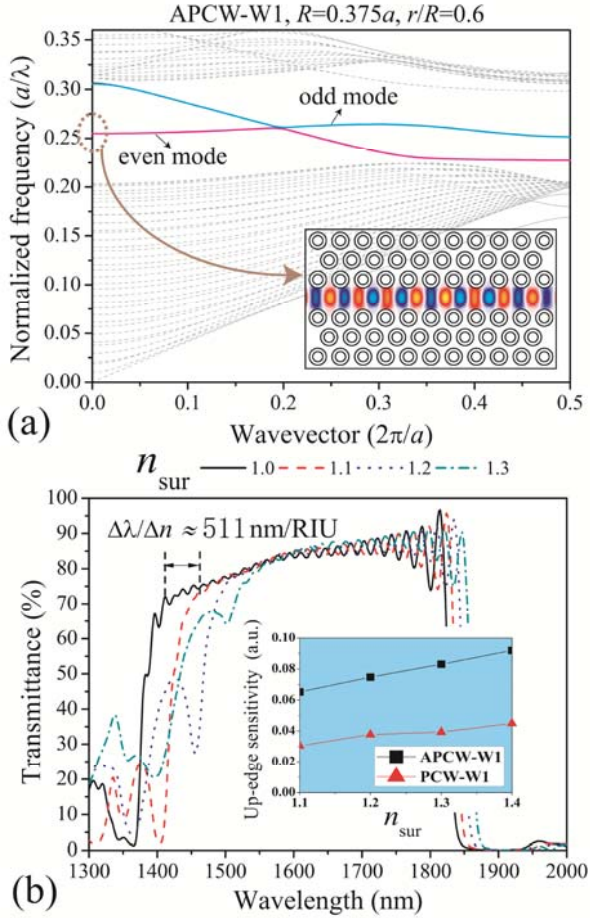
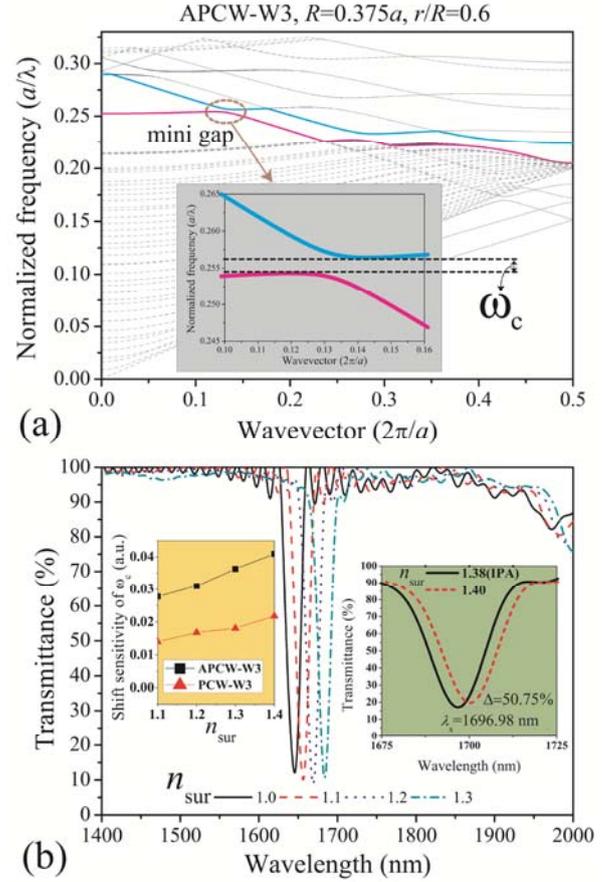

Fig. 5. (a) Project band diagram of the APCW-W1 when $R=0.375a$ and $r/R=0.6$. The inset shows the structure of APCW-W1 and the electric-filed profile of fundamental even mode. (b) Typical response of transmission spectra to surrounding-RI change for the APCW-W1. The inset shows the comparison of up-edge sensitivity of odd mode between APCW-W1 and conventional PCW-W1 with the same filling factor 32.65%.

The second typical structure of PCWs is PCW-W3, which is constructed by removing three lines of air holes in pure PCs along the propagating direction of electromagnetic waves. For such type of PCWs, there will be higher-order modes in the project band diagram. So there will be no novel physical mechanism if we still use the up frequency edge of the highest mode to make a RI sensing application, and it is also not suitable to do such application due to the higher transmission loss and unavoidable modes crosstalk during higher-order modes. Recent study [16] shows that the PCW-W3 can own a sub mini gap between the fundamental even and odd modes. Such mini gap can be used for RI sensing application because it can bring a 'dig-hole' effect in the transmission spectrum. Based on this idea, we made a further investigation on the APCW-W3 structure by studying the shift of central normalized frequency $\omega_c$ of the mini gap. As shown in Fig. 6(b), the APCW-W3 can produce about two times higher shifting sensitivity of $\omega_c$ than conventional PCW-W3 with the same filling factor 32.65%. However, it is not suitable to directly use such wavelength shift for RI sensing since the sensitivity is a bit lower than pure APCs or APC-W1. Instead of this, we can use the intensity shift at a particular wavelength since there is an apparent sharp dip in the transmission spectrum [seen in Fig. 6(b)]. This method is quite useful for acquiring very high detecting limitation than conventional wavelength-shift method. For instance, if we change the RI of surrounding media from 1.38 [Iso Propyle Alcohol (IPA) solution] to 1.40, there will be a 50.75% intensity change at the central wavelength 1696.98 nm for IPA solution. Consequently, an estimated RI detecting limitation of ~$3.9\times10^{-5}$ RIU can be reached by this method if the maximum output power is 1 mW and the power meter has a sensitivity of $10^{-3}$ mW.

Fig. 6. (a) Project band diagram of the APCW-W3 when $R=0.375a$ and $r/R=0.6$. The inset shows the zoomed mini gap with central normalized frequency $\omega_c$ between the fundamental even and odd modes. (b) Typical response of transmission spectra to surrounding-RI change for the APCW-W3. The left inset shows the comparison of shift sensitivity of $\omega_c$ between APCW-W3 and conventional PCW-W3 with the same filling factor 32.65%. The right inset shows the zoomed transmission spectra of APCW-W3 when surrounding RI varies from 1.38 (IPA) to 1.40.

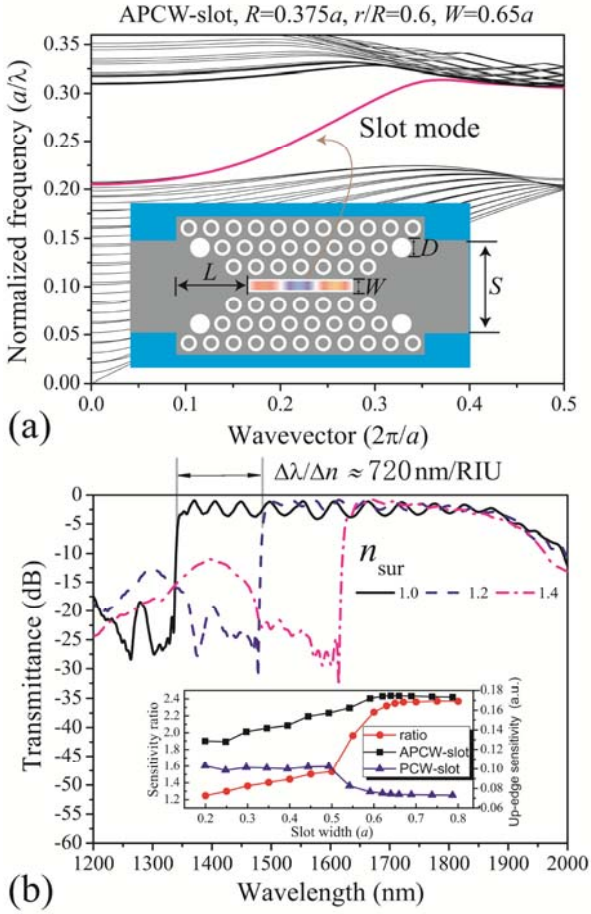

Fig. 7. (a) Project band diagram of the APCW-slot when $R=0.375a$, $r/R=0.6$ and air slot width $W$ is $0.65a$. The inset shows the structure of APCW-slot and the electric-filed profile of fundamental slot mode. $S$ and $L$ represent the width of accession waveguides and the distance of air slot to accession waveguides, while $D$ is the diameter of air-hole defect. (b) Typical response of transmission spectra to surrounding-RI change for the APCW-slot. The inset shows the comparison of up-edge sensitivity of slot mode between APCW-slot and conventional PCW-slot with the same filling factor 32.65% when slot width varies.

The final typical structure of PCWs is PCW-slot, which is constructed by replacing one line of air holes with an air slot in pure PCs along the propagating direction of electromagnetic waves. Recently, such type of PCWs has been extensively studied since it has several advantages compared with conventional PCW-W1 waveguides. The key advantage of such type of PCWs is its strong confinement of light in air that can promote strong light-matter interactions with a substance of interest. The air slot can conveniently been filled with analytes of interest, which can be readily sensed by the optical field. Another important advantage is its single mode feature in the project band diagram. As shown in Fig. 7(a), for an APCW-slot waveguide with filling factor 32.65%, there is only a fundamental slot mode during the PBG. Such single-mode property is very helpful for RI sensing because it can avoid the crosstalk between different modes. On the other hand, due to impendence mismatch, there will be a bit more coupling loss at the Silicon/air-slot interface. To decrease the coupling loss, a useful approach is optimizing the width of air slot ($W$) as well as the accession waveguides ($S$), the distance between air slot to accession waveguides ($L$), and modifying the configuration of PCW-slot at both input and output surfaces. As illustrated in the Fig. 7(a), we can modify the input and output surfaces of an typical APCW-slot by removing 8 air-ring units and replacing 4 air-ring units by pure air holes with a larger diameter $D$. The final optimal design of such APCW-slot is: $W=0.65a$, $S=4.49a$, $L=2.86a$, $D=0.45a$. Figure 7(b) shows the corresponding transmission spectra of this APCW-slot when surrounding RI varies from 1.0 to 1.4. We can observe clear cutoff in the transmission spectra (attributed to single slot mode feature) and a high RI sensitivity of ~720 nm/RIU. Compared with APCW-W1 waveguide, such high RI sensitivity in APCW-slot is attributed to the combination of APCW and air-slot waveguide. From the inset of Fig. 7(b), it can be clear found that the RI sensitivity of APCW-slot is influenced by the slot width. In particular, high RI sensitivity can be obtained when slot width is between $0.6a$ and $0.8a$ while the conventional PCW-slot waveguide will show an inverse behavior in this range. As a result, a large and robust sensitivity ratio of about 2.3 can be obtained in this range.

## 3. Dual–polarization APCs biosensors

It is well-known for 2D PCs that the air holes in the dielectric substrate highly favor the TE mode and the dielectric rods in the air medium highly favor the TM mode. In most of previous woks [11-18], studies on 2D PCs biosensors were focused on air-hole type and only TE polarization was considered. Compared with conventional 2D PCs, 2D APCs can well support both TE and TM modes because they are actually a combination of air-hole type and dielectric-rod type PCs. Based on such natural dual–polarization feature, we will present a new conception of dual–polarization RI biosensors by using pure 2D APCs.

Figures 8(a1) and 8(a2) show the TE and TM

band diagrams of a conventional 2D PC (in case of $R=0.45a$ and $r/R=0$) and a 2D APC (in case of $R=0.45a$ and $r/R=0.3$), respectively. For conventional 2D PC, there is a small TM PBG (between 2nd and 3rd band) embedded in TE PBG (between 1st and 2nd band), while for 2D APC there is a TM PBG (between 3rd and 4th band) with overlapped area with TE PBG (between 1st and 2nd band) located at lower frequency area. As mentioned in section 2.1, the relative RI sensitivity of pure 2D PCs can be represented by up-edge shift of either TE or TM PBG. We calculated the corresponding relative RI sensitivity for both TE and TM polarizations as illustrated in the left inset of Fig. 8(b). It can be found that TE polarization can show higher RI sensitivity than TM polarization for both conventional 2D PCs and 2D APCs, but the 2D PCs does not well support TM polarization compared with 2D APCs. As surrounding RI is increased to 1.3, the RI sensitivity will suffer an earlier termination due to the close of TM PBG. As a result, it is difficult to develop dual–polarization RI biosensors from conventional 2D PCs.

Now, let's transfer our attention to 2D APCs. Here we propose a new conception of dual–polarization RI biosensors by defining the relative sensitivity as following equation:

$$S_{(i+1)} = (1/\omega_{\_TE_{up\text{-}edge}}^{n_{sur(i+1)}} - 1/\omega_{\_TM_{up\text{-}edge}}^{n_{sur(i)}})$$
$$- (1/\omega_{\_TE_{up\text{-}edge}}^{n_{sur(i)}} - 1/\omega_{\_TM_{up\text{-}edge}}^{n_{sur(i+1)}})$$
$$= S_{(i+1)}^{TE \leftrightarrow TM} - S_{(i+1)}^{TM \leftrightarrow TE}$$

(2)

Here, $\omega_{\_TE_{up\text{-}edge}}^{n_{sur(i)}}$ and $\omega_{\_TM_{up\text{-}edge}}^{n_{sur(i)}}$ represent the up-edge frequency of TE and TM PBG when the $i$-th surrounding RI $n_{sur(i)}$ is considered. $S_{(i+1)}^{TE \leftrightarrow TM}$ and $S_{(i+1)}^{TM \leftrightarrow TE}$ represent the crossing relative sensitivity of up-edge PBG between TE and TM polarizations when surrounding RI varies from $n_{sur(i)}$ to $n_{sur(i+1)}$. In fact, equation (2) can be further written as:

$$S_{(i+1)} = (1/\omega_{\_TE_{up\text{-}edge}}^{n_{sur(i+1)}} - 1/\omega_{\_TE_{up\text{-}edge}}^{n_{sur(i)}})$$
$$+ (1/\omega_{\_TM_{up\text{-}edge}}^{n_{sur(i+1)}} - 1/\omega_{\_TM_{up\text{-}edge}}^{n_{sur(i)}})$$
$$= S_{(i+1)}^{TE} + S_{(i+1)}^{TM}$$

(3)

Here $S_{(i+1)}^{TE}$ and $S_{(i+1)}^{TM}$ represent the relative sensitivity of TE and TM polarizations according to equation (1).

Based on such definition, we can find that the relative RI sensitivity of 2D APCs will be equal to the sum of relative sensitivity for TE and TM polarizations rather than the relative sensitivity for mono polarization. Moreover, for a specific 2D APC such dual-polarization working mechanism will not only show higher relative sensitivity but also own other two advantages that could not be found in mono-polarization type. One advantage is that it will make the comparison of up-edge shift of PBG more convenient due to apparent separation of up-edge PBG between TE and TM polarizations when slight RI of ambient varies. Another critical advantage is that it will show more steady relative sensitivity when surrounding RI varies. As illustrated in the right inset of Fig. 8(b), $S_{(i+1)}^{TE \leftrightarrow TM}$ and $S_{(i+1)}^{TM \leftrightarrow TE}$ will show nearly parallel linear response to RI change that can result in steady difference between them. Compared with mono-polarization type [seen in Fig. 2 and Fig. 3], this advantage in fact can make the RI sensitivity suffer less influence from the close of PBG. We have made more calculations by comparing the dual-polarization relative sensitivity for different 2D APCs when $R$, $r/R$, or filling factor of air rings is fixed respectively. In particular, when $R$ is fixed [seen in Fig. 8(b)] 2D APCs with larger $r/R$ will show higher relative sensitivity in most area of RI detection range, while when $r/R$ or filling factor is fixed [seen in Fig. 8(c)] 2D APCs with relatively large $R$ can support higher dual-polarization RI sensitivity. Finally, to verify the conception of dual–polarization RI biosensors, we give a specific FDTD simulation of the reflection spectra for a 2D APC when $R=0.47a$ ($a=540$ nm) and air-ring filling factor is 61.71%. As shown in Fig. 8(d), when surrounding RI varies from 1.3 to 1.4 it is easy to find a cut-off wavelength difference of ~458 nm between $TE_{1.3}$ and $TM_{1.4}$ and a cut-off wavelength difference of ~577 nm between $TE_{1.4}$ and $TM_{1.3}$. This will bring a dual–polarization RI sensitivity of ~1190 nm/RIU in near-infrared area.

## 4. Conclusion

In summary, we have systematically investigated the RI sensitivity of 2D APCs. The results show that 2D APCs can support apparently enhanced RI sensitivity compare with conventional 2D PCs. For instance, compare with conventional 2D PCs with air-hole filling factor of 32.65% (in case of $R=0.3a$), the pure 2D APCs with the same air-ring filling factor can show as high as up to 2~3 times RI sensitivity. This attractive feature of 2D APCs is attributed to the suppressed up frequency edge of PBG when additional dielectric rods are introduced into air holes. Based on a specific 2D APC with $R=0.375a$ and $r/R=0.6$, we have further investigated three typical waveguides, that is APCW-W1, APCW-W3, and APCW-W1, APCW-slot to develop sorts of possible

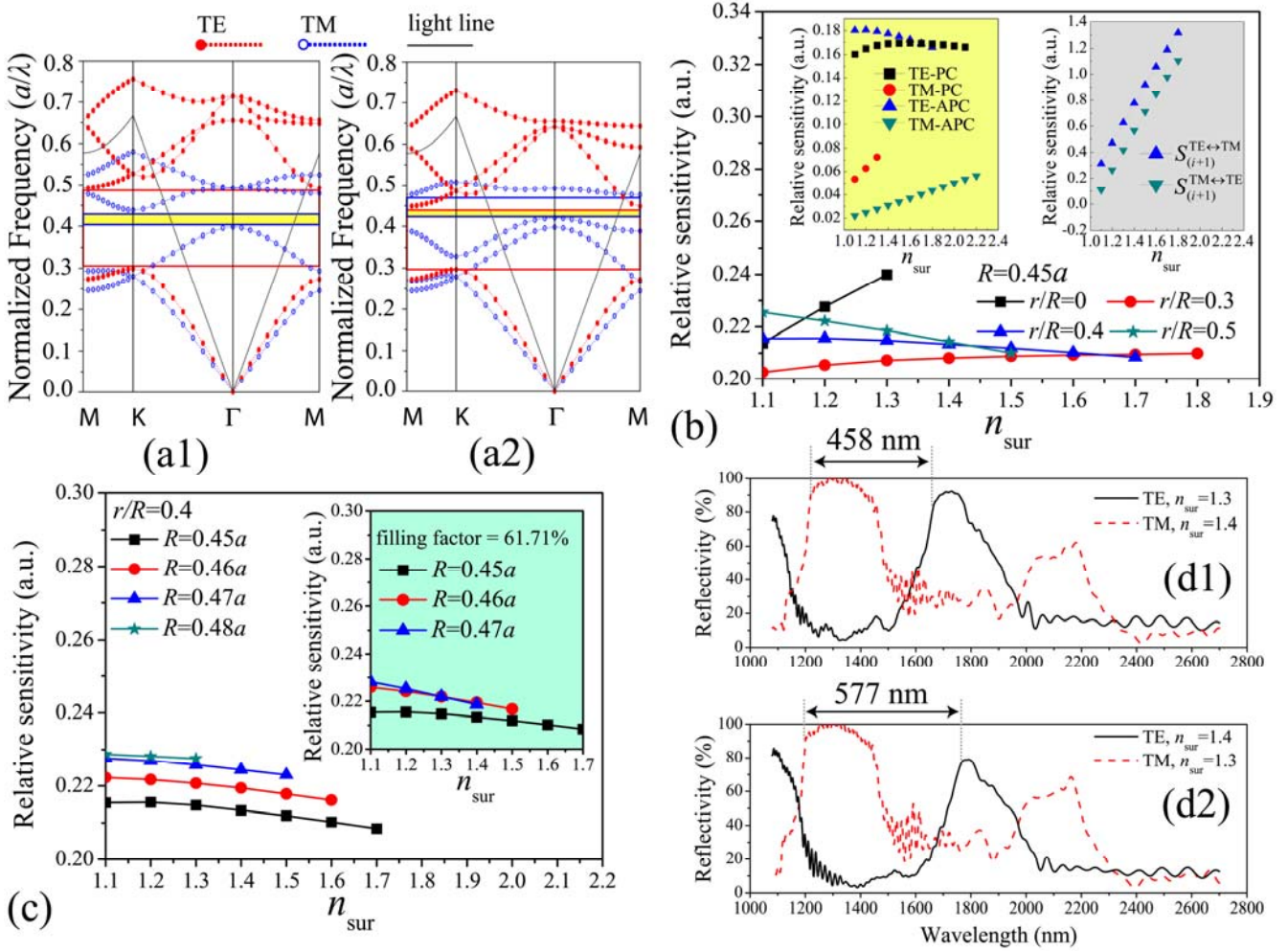

Fig. 8. (a1) and (a2) Comparison of band diagram between pure 2D PC ($R=0.45a$) and 2D APC ($R=0.45a$ and $r/R=0.3$) for both TE and TM polarizations. (b) The response of dual-polarization relative RI sensitivity to surrounding-RI change for different 2D APCs with fixed $R$. The left inset shows the comparison of relative sensitivity for TE and TM polarizations between 2D PC ($R=0.45a$) and 2D APC ($R=0.45a$ and $r/R=0.3$). The right inset shows the crossing relative sensitivity of up-edge PBG between TE and TM polarizations for pure 2D APC ($R=0.45a$ and $r/R=0.3$). (c) The response of dual-polarization relative RI sensitivity to surrounding-RI change for different 2D APCs when fixed $r/R$ or air-ring filling factor (see inset) is considered. (d) Typical reflection spectra for pure 2D APC ($R=0.47a$ and air-ring filling factor is 61.71%) when surrounding RI changes from 1.3 to 1.4.

APCs RI biosensors with satisfactory performance. In particular, the APCW-W3 can exhibit a quite low RI detecting limitation of ~$3.9\times10^{-5}$ RIU while the APCW-slot can show a high RI sensitivity of ~720 nm/RIU for bio-sensing applications in infrared range. Moreover, we have proposed a new conception of dual–polarization RI biosensors by calculating the difference of crossing relative sensitivity of up-edge PBG between TE and TM polarizations. Such dual–polarization RI biosensors can show an enhanced relative sensitivity that is in fact equal to the sum of relative sensitivity for two polarizations. A dual–polarization RI biosensor based on pure 2D APCs was finally verified to show a high RI sensitivity of ~1190 nm/RIU in infrared range. Besides pure PCs and PCWs, the physical concept of this work can also be easily extended to PCs cavities, such as 1D [9] or 2D [16-18] PCs microcavities developed from APCs. It is strongly encouraged that other groups in related with biosensors or chemical sensors can do further experiments to demonstrate the enhanced RI sensitivity in APCs-based micro/nano devices.

## Acknowledgement

This work was supported by the Natural Science Foundation of China (No: 61205042) and the Natural Science Foundation of Jiangsu Province in China (No: BK2014021828). The first author also wishes to acknowledge the financial support from the China Scholarship Council as well as the Zijin Intelligent Program

**References**

[1] X.D. Fan, I.M. White, S.I. Shopova, H.Y. Zhu, J.D. Suter, Y.Z. Sun, Sensitive optical biosensors for unlabeled targets: A review, Anal. Chim. Acta 620 (2008) 8-26.

[2] A.J. Haes, R.P. Van Duyne, A unified view of propagating and localized surface plasmon resonance biosensors, Anal. Bioanal. Chem. 379 (2004) 920-930.

[3] S.Y. Gao, N. Koshizaki, H. Tokuhisa, E. Koyama, T. Sasaki, J.K. Kim, et al., Highly Stable Au Nanoparticles with Tunable Spacing and Their Potential Application in Surface Plasmon Resonance Biosensors, Adv. Funct. Mater. 20 (2010) 78-86.

[4] F.F. Pang, H.H. Liu, H.R. Guo, Y.Q. Liu, X.L. Zeng, N. Chen, et al., In-Fiber Mach-Zehnder Interferometer Based on Double Cladding Fibers for Refractive Index Sensor, IEEE Sens. J. 11 (2011) 2395-2400.

[5] K. Schmitt, K. Oehse, G. Sulz, C. Hoffmann, Evanescent field sensors based on tantalum pentoxide waveguides - A review, Sensors 8 (2008) 711-738.

[6] F. Vollmer, S. Arnold, Whispering-gallery-mode biosensing: label-free detection down to single molecules, Nat. Methods 5 (2008) 591-596.

[7] A. Leung, P.M. Shankar, R. Mutharasan, A review of fiber-optic biosensors, Sens. Actuators B: Chem. 125 (2007) 688-703.

[8] R.V. Nair, R. Vijaya, Photonic crystal sensors: An overview, Prog. Quantum Electron. 34 (2010) 89-134.

[9] S. Mandal, J.M. Goddard, D. Erickson, A multiplexed optofluidic biomolecular sensor for low mass detection, Lab Chip 9 (2009) 2924-2932.

[10] V.N. Konopsky, E.V. Alieva, A biosensor based on photonic crystal surface waves with an independent registration of the liquid refractive index, Biosens. Bioelectron. 25 (2010) 1212-1216.

[11] M. El Beheiry, V. Liu, S.H. Fan, O. Levi, Sensitivity enhancement in photonic crystal slab biosensors, Opt. Express 18 (2010) 22702-22714.

[12] S.E. Baker, M.D. Pocha, A.S.P. Chang, D.J. Sirbuly, S. Cabrini, S.D. Dhuey, et al., Detection of bio-organism simulants using random binding on a defect-free photonic crystal, Appl. Phys. Lett. 97 (2010) 113701.

[13] M. Scullion, T. Krauss, A. Di Falco, Slotted Photonic Crystal Sensors, Sensors 13 (2013) 3675-3710.

[14] S. Pal, E. Guillermain, R. Sriram, B.L. Miller, P.M. Fauchet, Silicon photonic crystal nanocavity-coupled waveguides for error-corrected optical biosensing, Biosens. Bioelectron. 26 (2011) 4024-4031.

[15] M.G. Scullion, A. Di Falco, T.F. Krauss, Slotted photonic crystal cavities with integrated microfluidics for biosensing applications, Biosens. Bioelectron. 27 (2011) 101-105.

[16] L. Cao, Y.D. Huang, X.Y. Mao, F. Li, W. Zhang, J.D. Peng, Fluid sensor based on transmission dip caused by mini stop-band in photonic crystal slab Chin. Phys. Lett. 25 (2008) 2101-2103.

[17] D. Dorfner, T. Zabel, T. Hurlimann, N. Hauke, L. Frandsen, U. Rant, et al., Photonic crystal nanostructures for optical biosensing applications, Biosens. Bioelectron. 24 (2009) 3688-3692.

[18] S. Zlatanovic, L.W. Mirkarimi, M.M. Sigalas, M.A. Bynum, E. Chow, K.M. Robotti, et al., Photonic crystal microcavity sensor for ultracompact monitoring of reaction kinetics and protein concentration, Sens. Actuators B: Chem. 141 (2009) 13-19.

[19] J. Wu, D. Day, M. Gu, A microfluidic refractive index sensor based on an integrated three-dimensional photonic crystal, Appl. Phys. Lett. 92 (2008) 071108.

[20] Y.N. Zhu, Z.H. He, H. Du, Detection of external refractive index change with high sensitivity using long-period gratings in photonic crystal fiber, Sens. Actuators B: Chem. 131 (2008) 265-269.

[21] X.D. Fan, I.M. White, Optofluidic microsystems for chemical and biological analysis, Nat. Photonics 5 (2011) 591-597.

[22] H. Kurt, D.S. Citrin, Annular photonic crystals, Opt. Express 13 (2005) 10316-10326.

[23] A. Cicek, B. Ulug, Polarization-independent waveguiding with annular photonic crystals, Opt. Express 17 (2009) 18381-18386.

[24] L.Y. Jiang, H. Wu, W. Jia, X.Y. Li, Polarization-independent negative refraction effect in SiO2-GaAs annular photonic crystals, J. Appl. Phys. 111 (2012) 023508.

[25] H. Wu, D.S. Citrin, L.Y. Jiang, X.Y. Li, Polarization-independent slow light in annular photonic crystals, Appl. Phys. Lett. 102 (2013) 141112.

[26] A. Saynatjoki, M. Mulot, K. Vynck, D. Cassagne, J. Ahopelto, H. Lipsanen, Properties, applications and fabrication of photonic crystals with ring-shaped holes in silicon-on-insulator, Photonics Nanostruct. 6 (2008) 42-46.

[27] H. Kurt, M.N. Erim, N. Erim, Various photonic crystal bio-sensor configurations based on optical surface modes, Sens. Actuators B: Chem. 165 (2012) 68-75.